\documentclass[aps,prd,showpacs,preprintnumbers,twocolumn,groupedaddress,nofootinbib]{revtex4-1}

\usepackage[]{amssymb}
\usepackage[]{amsmath}
\usepackage[]{mathrsfs}
\usepackage[]{eucal}
\usepackage[]{tensor}
\usepackage[]{amsthm}
\usepackage[]{bm}

\newcommand{\dd}{\partial}
\newcommand{\df}{\mathrm{d}}
\newcommand{\w}{\wedge}
\newcommand{\Lie}{\pounds}
\newcommand{\nab}[1]{\nabla_{\!#1}}
\newcommand{\qqd}{\ , \quad}
\newcommand{\bc}{\begin{center}}
\newcommand{\ec}{\end{center}}
\newcommand{\be}{\begin{equation}}
\newcommand{\ee}{\end{equation}}

\newcommand{\FF}{\mathcal{F}}
\newcommand{\GG}{\mathcal{G}}
\newcommand{\LL}{\mathscr{L}}

\newcommand{\nn}{\mathbb{N}}

\newcommand{\cl}[1]{\overline{#1}}

\usepackage[unicode]{hyperref}
\usepackage{xcolor}
\definecolor{pastgreen}{HTML}{669900}
\definecolor{pastblue}{HTML}{336699}
\definecolor{linkcol}{HTML}{663333}
\hypersetup{colorlinks,linkcolor={pastblue},citecolor={pastblue},urlcolor={pastblue}}

\theoremstyle{plain} \newtheorem{tm}{Theorem}[section]
\theoremstyle{plain} \newtheorem{lm}[tm]{Lemma}
\theoremstyle{plain} \newtheorem{cor}[tm]{Corollary}
\theoremstyle{definition} \newtheorem{defn}[tm]{Definition}
\theoremstyle{definition} 
\newcommand{\btm}{\begin{tm}}
\newcommand{\etm}{\end{tm}}
\newcommand{\blm}{\begin{lm}}
\newcommand{\elm}{\end{lm}}
\newcommand{\bcor}{\begin{cor}}
\newcommand{\ecor}{\end{cor}}
\newcommand{\bdefn}{\begin{defn}}
\newcommand{\edefn}{\end{defn}}

\begin{document}

\preprint{ZTF-EP-17-11}

\title{Spacetimes dressed with stealth electromagnetic fields}

\author{Ivica Smoli\'c}
\email[]{ismolic@phy.hr}

\affiliation{Department of Physics, Faculty of Science, University of Zagreb, 10000 Zagreb, Croatia}

\date{\today}

\begin{abstract}
Stealth field configurations by definition have a vanishing energy-momentum tensor, and thus do not contribute to the gravitational field equations. While only trivial fields can be stealth in Maxwell's electrodynamics, nontrivial stealth fields appear in some nonlinear models of electromagnetism. We find the necessary and sufficient conditions for the electromagnetic fields to be stealth and analyse which models admit such configurations. Furthermore, we present some concrete exact solutions, featuring a class of black holes dressed with the stealth electromagnetic hair, closely related to force-free solutions. Stealth hair does not alter the generalized Smarr formula, but may contribute to the Komar charges.
\end{abstract}


\maketitle

\section{Introduction}

Gravitational field equations explicate exactly how matter and gauge fields curve spacetime. One might then naively expect that any nonvanishing field will inevitably leave its imprint on the spacetime it inhabits. However, as was noticed a decade ago \cite{ABMTZ05} (see also the analysis in \cite{Sokol03}), some exact solutions of the vacuum Einstein's equations are simultaneously exact solutions of the Einstein-Klein-Gordon equations with nontrivial, nonminimally coupled, real scalar fields. In other words, it is possible to have seemingly nongravitating field configurations, which were aptly dubbed \emph{stealth} fields. Stealth scalar fields have been found on top of the BTZ black hole \cite{ABMZ06}, four-dimensional black holes \cite{BC14} and some cosmological solutions \cite{BJJ08,AGRT13,BGH13a,BGH13b,CCH16}. Stability of stealth configurations was analysed in \cite{FM10} and their classification, from the perspective of the symmetry inheritance, was presented in \cite{BS17}. Such solutions have been also found in Brans-Dicke-Maxwell theory \cite{Robinson06} and, more recently, within various vector-tensor models \cite{CHOR16,HKMT17,Fan17,CT17}.

\medskip

What about the \emph{minimally} coupled electrodynamic fields? As is shown below, nontrivial stealth configurations are absent in Maxwell's electrodynamics. There is, however, a large class of \emph{nonlinear} electrodynamic models, which extend the canonical linear one. Nonlinear electrodynamics (NLE) appears in the quantum corrections to the classical theory, either in quantum electrodynamics \cite{HE36} or in low energy limits of the string theory \cite{FT85,SW99}, and can solve the inconsistencies in the classical theory related to the point charges \cite{BI34,HE36} and the spacetime singularities \cite{deO94,ABG98,ABG99,ABG00,Bronnikov00,GSB00,CGMCL04,FW16}. Any nonlinear electromagnetic stealth solution would be an example of a classical configuration ``resistant'' to some quantum corrections.

\medskip

Limits on the parameters of the nonlinear electromagnetic models have been placed by a series of experiments, the most important of which are focused on the vacuum birefringence and the photon-photon scattering \cite{BR13,FBR16}. A recent strong constraint on the mass scale of the Born-Infeld model (whose lower bound is now placed above 100 GeV) has been inferred from the results of the measurement of the ATLAS collaboration \cite{EMY17}. Future experiments will include new generations of the ultraintense lasers \cite{DiPMHK12,ELI} and the astrophysical tests \cite{Mig17}.

\medskip

Before we proceed, let us make several brief remarks on notation and conventions. The metric signature is always $(-,+,+,+)$ and the system of units is a natural one with $c = G = 4\pi\varepsilon_0 = 1$. We use both the abstract index notation (see e.g.~\cite{Wald}) and the ``indexless'' notation (see e.g.~\cite{Heusler}), the former at places where we want to emphasize the type of the tensor and the latter whenever the abstract indices may be suppressed in order to avoid cumbersome expressions. The Hodge dual of a $p$-form $\omega_{a_1 \dots a_p}$ is denoted by a star and defined as
\be
{*\omega}_{b_1 \dots b_{4-p}} = \frac{1}{p!}\,\omega_{a_1 \dots a_p} \tensor{\epsilon}{^{a_1}^{\dots}^{a_p}_{b_1}_{\dots}_{b_{4-p}}} \ ,
\ee
where $\epsilon_{abcd}$ is the Levi-Civita tensor. For example, the volume form may be compactly written as ${*1}$.

\medskip

We consider a general class of four-dimensional models of NLE, described by the Lagrangians of the form
\be\label{eq:LagNLE}
L_{\mathrm{EM}} = \LL(\FF,\GG)\,{*1} \ ,
\ee
where we have introduced two standard electro\-mag\-netic invariants, 
\be
\FF \equiv F_{ab} F^{ab} \qquad \mathrm{and} \qquad \GG \equiv F_{ab}\,{*F}^{ab} \ .
\ee
For example, canonical Maxwell's Lagrangian is
\be\label{eq:LMax}
L_{\mathrm{EM}}^{\mathrm{(Max)}} = -\frac{1}{4}\,\FF\,{*1} \ .
\ee
The energy-mo\-men\-tum tensor corresponding to (\ref{eq:LagNLE}) may be written in a convenient way,
\be\label{eq:TTMax}
T_{ab} = - 4\LL_\FF\,T_{ab}^{\mathrm{(Max)}} + \frac{1}{4}\,T g_{ab} \ ,
\ee
with the help of the abbreviation $\LL_\FF \equiv \dd\LL/\dd\FF$, the trace
\be
T \equiv g^{ab} T_{ab} = \frac{1}{\pi} \left( \LL - \LL_\FF \FF - \LL_\GG \GG \right)
\ee
and Maxwell's energy-mo\-men\-tum tensor,
\be\label{eq:TEM}
T_{ab}^{\mathrm{(Max)}} = \frac{1}{4\pi} \left( F_{ac} \tensor{F}{_b^c} - \frac{1}{4}\,g_{ab}\,\FF \right) \ .
\ee
Finally, the set of generalized gravitational-non\-linear Max\-well's (gNLM) field equations consists of
\be\label{eq:EOM}
E_{ab} = 8\pi T_{ab} \qqd \df F = 0 \quad \textrm{and} \quad \df\,{*Z} = 0 \ .
\ee
The symmetric tensor $E_{ab}$ is any diff-invariant gravitational tensor (e.g.~Einstein's tensor $G_{ab}$), constructed out of the spacetime metric, its derivatives and possibly the Levi-Civita tensor. The auxiliary 2-form
\be
Z_{ab} = -4(\LL_\FF\,F_{ab} + \LL_\GG\,{*F}_{ab})
\ee 
is defined with $\LL_\GG \equiv \dd\LL/\dd\GG$.

\section{How to hide the electromagnetic field}

Let us start with the formal definition of the central object of this discussion.

\smallskip

\bdefn
We say that a nonzero electromagnetic field $F_{ab}$ is \emph{stealth} if the corresponding energy-momentum tensor identically vanishes, $T_{ab} = 0$.
\edefn

\smallskip

The easiest way to see that only trivial fields can be stealth in Maxwell's electrodynamics is to write the energy-momentum tensor in the spinorial formalism \cite{Stewart},
\be
T_{ABA'B'}^{\mathrm{(Max)}} = -\frac{1}{2\pi}\,\phi_{AB}\cl{\phi}_{A'B'} \ .
\ee
If the field is nontrivial, $\phi_{AB} \ne 0$, then there are spinors $\alpha^A$ and $\beta^B$, such that $\phi_{AB} \alpha^A \beta^B \ne 0$. But then the contraction of the stealth condition
\be
T_{ABA'B'}^{\mathrm{(Max)}} = 0
\ee
with $\overline{\alpha}^{A'}\overline{\beta}^{B'}$ implies $\phi_{AB} = 0$, a contradiction. Therefore, the linear electromagnetic field is stealth if and only if it is trivial. For a general class of NLE fields we have the following characterisation of the stealth configurations.

\smallskip

\btm\label{tm}
Suppose that NLE field is nontrivial, $F_{ab} \ne 0$. Then this field is stealth if and only if both $T = 0$ and $\LL_\FF = 0$ are satisfied. Furthermore, a stealth NLE field solution of (\ref{eq:EOM}) at each point where $\df\LL_\GG \ne 0$ holds necessarily satisfies $\GG = 0$.
\etm

\smallskip

Stealth condition $T_{ab} = 0$ implies $T = 0$ and, consequently, $\LL_\FF\,T_{ab}^{\mathrm{(Max)}} = 0$. However, as has been shown above, $T_{ab}^{\mathrm{(Max)}} = 0$ holds if and only if $F_{ab} = 0$. Thus, it follows that $\LL_\FF = 0$. The converse is trivial. Finally, the second generalized Maxwell's equation for the stealth field becomes
\be
\df \LL_\GG \w F = 0 \ ,
\ee
which is equivalent to the condition $(\nabla^a \LL_\GG)\,{*F}_{ab} = 0$. Hence, at each point where $\nabla^a \LL_\GG \ne 0$ the field ${*F}_{ab}$ is necessarily simple and degenerate \cite{GJ14}, so that $\GG = 0$. If we have a more special class of NLE models with $\LL = \LL(\FF)$, then the stealth fields automatically satisfy the second Maxwell's equation, while the invariant $\GG$ does not necessarily vanish.

\medskip

Note that in any NLE model defined by a Lagrangian which respects the proper Maxwell's asymptotics in the weak field limit ($\LL \to 0$ and $\LL_\FF \to -1/4$ as $\FF \to 0$), the stealth field \emph{cannot be null}.

\subsection{NLE models}

A number of NLE models that have been extensively studied in the literature do not admit stealth configurations, simply because $\LL_\FF \ne 0$ for any real field $F_{ab}$. Among these we have Born-Infeld \cite{BI34}, Bardeen \cite{Bardeen68,ABG00}, Soleng's logarithmic \cite{Soleng95}, Hendi's exponential \cite{Hendi12,Hendi13} or Kruglov's rational \cite{Kruglov15} and arcsin Lagrangian \cite{Kruglov16}.

\medskip

On the other hand, the power-Maxwell model \cite{HM07,HM08} defined with $\LL = C\FF^s$, where $C \ne 0$ and $s > 1$ are some real constants, admits stealth configurations. In order to avoid some unphysical solutions \cite{HM08} we must choose parameter $s$ to be a rational number which written in lowest terms has an odd denominator, $s = m/(2n-1)$ with $m,n\in\nn$. As long as $s > 1$, the necessary and sufficient stealth condition is that $\FF = 0$. Similarly, the Hoffmann-Infeld model \cite{HI37,AFG05} admits stealth configurations in the limit when $\FF \to 0$.

\medskip

Finally, let us look at two classes of weak-field limit Lagrangians which satisfy proper Maxwellian asymptotics. First, if $\LL = \LL(\FF)$ is a smooth function on some neighbourhood of $\FF = 0$, in the weak-field limit it has a form 
\be\label{eq:modelF}
\LL = -\frac{1}{4}\,\FF + \alpha\FF^2 + \beta\FF^3
\ee
with some real constants $\alpha$ and $\beta$. Then the stealth conditions from the theorem \ref{tm} imply $\beta = -\alpha^2$ and $\FF = (2\alpha)^{-1}$.

\medskip

The other class of models is defined by the Euler-Heisenberg type of Lagrangians \cite{HE36,Kruglov17}, 
\be\label{eq:modelFG}
\LL = -\frac{1}{4}\,\FF + \kappa\FF^2 + \lambda\GG^2 \ ,
\ee
again with some real constants $\kappa$ and $\lambda$. Here the stealth conditions from the theorem \ref{tm} imply $\FF = (8\kappa)^{-1}$ and $\GG^2 = -(64\kappa\lambda)^{-1}$ (since $\GG$ is constant we immediately have $\df\LL_\GG = 0$). Whence, the necessary condition for such models to admit stealth configuration is that $\kappa\lambda < 0$. This immediately eliminates the original Euler-Heisenberg's Lagrangian \cite{HE36} since there we have quite the opposite, $\kappa\lambda > 0$.

\subsection{Black hole solutions}

One particularly important perspective of the stealth configurations is to provide novel black hole hair. The expression for the Komar electric charge $Q_{\mathcal{S}}$ \cite{Komar63,JG11,GS17} of a stealth configuration, contained in a closed smooth 2-surface $\mathcal{S}$, is reduced to the integral
\be
Q_{\mathcal{S}} = \frac{1}{4\pi} \int_{\mathcal{S}} {*Z} = -\frac{1}{\pi} \int_{\mathcal{S}} \LL_\GG \, F \ .
\ee
So, if $\LL = \LL(\FF)$, the electric charge identically vanishes, while in a more general class of models (\ref{eq:LagNLE}) it may be nonzero. We note in passing that the stealth conditions do not impose any direct constraint on the Komar magnetic charge. 

\medskip

One of the central relations in black hole thermodynamics, the Smarr formula \cite{Smarr72}, has been recently generalized for any stationary axially symmetric black hole. Technically, we assume that the spacetime is stationary, axially symmetric, asymptotically flat and contains a connected Killing horizon $H[\chi]$, generated by the Killing vector field $\chi^a = k^a + \Omega_{\mathsf{H}} m^a$, where $k^a$ is the stationary and $m^a$ is the axial Killing vector field. Then, in the presence of symmetry inheriting NLE fields \cite{GS17}, the generalized Smarr formula reads
\be
M = \frac{\kappa\mathcal{A}}{4\pi} + 2\Omega_{\mathsf{H}} J + \Phi_{\mathsf{H}} Q_{\mathsf{H}} + \Psi_{\mathsf{H}} P_{\mathsf{H}} + \Delta \ .
\ee
Here we have Komar mass $M$, surface gravity $\kappa$, horizon area $\mathcal{A}$, angular velocity of the horizon $\Omega_{\mathsf{H}}$, Komar angular momentum $J$, electrostatic potential at the horizon $\Phi_{\mathsf{H}}$ (written in gauge in which $\Phi \to 0$ at infinity), Komar electric charge $Q_{\mathsf{H}}$, magnetostatic potential at the horizon $\Psi_{\mathsf{H}}$ (written in gauge in which $\Psi \to 0$ at infinity) and Komar magnetic charge $P_{\mathsf{H}}$. The additional term $\Delta$ is given by the integral 
\be
\Delta = \frac{1}{2} \int_\Sigma T\,{*\chi}
\ee
over a smooth spacelike hypersurface $\Sigma$ which intersects the black hole horizon. Since the trace $T$ for the stealth electromagnetic fields identically vanishes, we necessarily have $\Delta = 0$, so that the Smarr formula remains unaltered. The stealth contributions to the Smarr formula vanish even if the NLE fields do not inherit the spacetimes symmetries, as can be seen from the Bardeen-Carter-Hawking mass formula (see e.g.~equation (25) in \cite{GS17}).

\section{Dressing the spacetimes with stealth electromagnetic fi\-elds}

From the discussion in the previous section we see that the stealth fields may be found among closed 2-forms $F_{ab}$ for which the invariants $\FF$ and $\GG$ are constant. The fact that the null electromagnetic fields (those in which $\FF$ and $\GG$ are both zero) are simultaneously solutions of a much wider class of electromagnetic field equations has been already recognized by Schr\"odinger \cite{S35} and recently reanalysed \cite{OP16,OP17} in the context of so-called universal solutions. Examples of the null electromagnetic fields can be relatively easily constructed as follows. Take any null vector field $\ell^a$, such that $\nab{[a}\ell_{b]} = 0$, and a function $\sigma$, such that $\ell^a \nab{a} \sigma = 0$. Then for $F_{ab} = \ell_{[a} \nab{b]}\sigma$ we have $\df F = 0$ and $\FF = 0 = \GG$. The non-null case demands more careful construction.

\subsection{Stealth from force-free electrodynamics}

Force-free electrodynamics, used for modeling of the magnetically dominated plasma around compact astrophysical objects, treats the solutions of the Maxwell's equations, $\df F = 0$ and $\df {*F} = 4\pi J$, under the assumption that $J^a F_{ab} = 0$. Some important force-free solutions are null \cite{BGJ13,GJ14} and may be ``recycled'' as a stealth NLE fields. Namely, we can find here examples of closed null fields $F_{ab}$, which immediately become a stealth field in any NLE model for which $\FF = 0$ (and $\GG = 0$) is a sufficient stealth condition. 

\medskip

Note two important differences: (1) whereas the force-free electromagnetic fields are just \emph{test} fields (solutions of the Maxwell's equations with nonvanishing current $J^a \ne 0$ on top of the fixed background spacetime), NLE counterparts are \emph{exact} solutions (of the source-free gNLE field equations (\ref{eq:EOM})), and (2) unlike in the force-free counterpart, any flux $T_{ab} X^a Y^b$ identically vanishes for the stealth solutions, so that stealth fields cannot be used to extract energy from the system.

\subsection{Minkowski's new clothes}

Suppose we want to find a NLE stealth field on top of the Minkowski spacetime. An example of a null electromagnetic field is given by $F = \df v \w \df\sigma$, where $\sigma = \sigma(v,y,z)$ and $v = t + x$ is a lightlike coordinate. For this field we immediately have $\df F = 0$ and, assuming that $\LL = \LL(\FF)$, $\df\,{*Z} = 0$ (this is an example of power-Maxwell NLE stealth field). The quantity $\sigma$ that appears in this solution can be directly related to the components of the electric and the magnetic field. For example, if the 4-velocity of a stationary observer is $u^a= \dd_t^a$, then the electric 1-form is given by $E = -i_u F = -\sigma_{,y}\,\df y -\sigma_{,z}\,\df z$ and the magnetic 1-form is $H = i_u {*Z} = 4 \LL_\FF (\sigma_{,z}\,\df y - \sigma_{,y}\,\df z)$.

\medskip

Another, more intriguing class of null electromagnetic fields can be found among the ``electromagnetic knots'' \cite{Ranada89,KBBPSI13}. These are the solutions of Maxwell's equations on a fixed Minkowski background, whose construction was based on the Hopf fibration of the 3-sphere. It was recently recognized \cite{Goulart16,AAHNS17} that the electromagnetic knots are simultaneously solutions in some NLE models, in accordance with the remarks from the beginning of this section. What we can add here is that these fields are also exact solutions of the gNLM field equations (\ref{eq:EOM}), as long as the NLE model is such that null electromagnetic fields satisfy the stealth conditions. An open question is how to use the Hopf fibration to construct knotted electromagnetic fields in a curved spacetime, an exact solution of the gravitational-electromagnetic field equations\footnote{A recently found ``Hopfionic'' solution of the Einstein-Maxwell field equations \cite{KN17} is a spacetime with $\mathbb{R} \times \mathbb{S}^3$ topology which, apart from the electromagnetic field, contains additional neutral matter}.

\medskip 

For a non-null Ansatz we may take a closed 2-form,
\be
F = -a\,\df t \w \df x + b\,\df y \w \df z
\ee
with some real constants $a$ and $b$. The electromagnetic invariants in this example are given by $\FF = 2(b^2 - a^2)$ and $\GG = 4ab$, so that their value, required by the stealth conditions in models (\ref{eq:modelF}) and (\ref{eq:modelFG}), can be fixed by an appropriate choice of the constants $a$ and $b$.

\subsection{Black holes with stealth hair}

A static, spherically symmetric spacetime metric written with ingoing Edding\-ton-Finkelstein coordinates $\{v,r,\theta,\varphi\}$, where $\df v = \df t + \df r_*$ and $\df r_* = \df r/f(r)$, is given by
\be\label{eq:sphg}
\df s^2 = -f(r)\,\df v^2 + 2 \df v \, \df r + r^2 (\df\theta^2 + \sin^2\theta\,\df\varphi^2) \ .
\ee
A class of null electromagnetic fields, stealth fields for the power-Maxwell NLE, can be constructed as above,
\be\label{eq:F1}
F = \df v \w \df\sigma \ ,
\ee
with $\sigma = \sigma(v,\theta,\varphi)$. The field (\ref{eq:F1}) is exactly one of the force-free solutions presented in \cite{GJ14}.

\medskip

This example demonstrates how NLE fields may evade some well-known no-go theorems. First, nontrivial null electromagnetic fields cannot exist in a static spacetime \cite{Tod06} and the extension of this theorem \cite{BGS17} holds in NLE models under the assumption that $\LL_\FF \ne 0$, which is broken by the stealth fields. Second, a linear electromagnetic field inherits symmetries in general spherically symmetric spacetime \cite{MW75,WY76a,WY76b}, but this does not necessarily hold in NLE models \cite{BGS17}. Although the points with vanishing $\LL_\FF$ are completely out of the scope of the analysis in \cite{BGS17}, the field (\ref{eq:F1}) nevertheless obeys the same constraints on breaking of the symmetry inheritance: for any Killing vector $K^a$ of the metric (\ref{eq:sphg}) the Lie derivative $\Lie_K F_{ab}$ is some linear combination of $F_{ab}$ and its Hodge dual ${*F}_{ab}$.

\medskip

Another class of stealth solutions for the model (\ref{eq:modelF}) is given by the 2-form
\be\label{eq:F2}
F = a\,\df v \w \df r + b\,\frac{r}{\sqrt{f(r)}}\,\df r \w \df \theta \ ,
\ee
with two real constants $a$ and $b$, for which $\FF = 2(b^2 - a^2)$ and $\GG = 0$. Note, however, that this solution is singular on the black hole horizon (defined by the condition $f(r) = 0$), unless $b = 0$, in which case we need $\alpha < 0$ in (\ref{eq:modelF}) for consistency with $\FF < 0$.  The corresponding electric and magnetic Komar charges for both black hole solutions (\ref{eq:F1}) and (\ref{eq:F2}), evaluated on any 2-surface defined by $v = \mathrm{const.}$ and $r = \mathrm{const.}$, are zero. 

\medskip

Finally, we can use a null force-free solution (see \cite{GJ14}, Secs.~4.4 and 4.5) on top of the Kerr black hole, written in Kerr ingoing coordinates, 
\be\label{eq:F3}
F = \df\sigma(\theta,\bar{\varphi}) \w (\df v - a \sin^2\theta\,\df\bar{\varphi}) \ ,
\ee
which is simultaneously a stealth solution in the power-Maxwell model. One distinguishing property of this solution is that it has a nonvanishing magnetic charge, which is proportional to the spin parameter $a$,
\be
P_\mathcal{S} = \frac{1}{4\pi} \int_{\mathcal{S}} F = -\frac{a}{4\pi}\,\int_\mathcal{S} \sigma_{,\theta} \sin^2\theta\,\df\theta \w \df\bar{\varphi} \ .
\ee
Again, $\mathcal{S}$ is a 2-surface defined by $v = \mathrm{const.}$ and $r = \mathrm{const.}$ This proves that it is possible in principle to have a black hole hair which, despite the fact that it is stealth, still contributes to the black hole charges. One might also remark that the freedom of choice of the function $\sigma$ qualifies this configuration as a primary hair \cite{CPW92}.

\section{Final remarks}

Like many other exotic theoretical constructions, stealth fields are the cornucopia of counterintuitive examples. Using a simple criterion, given in theorem \ref{tm}, we have proven that NLE models which admit such configurations are rare among those which are most often analysed in the literature. However, whereas the nontrivial stealth fields are absent in linear Maxwell's electrodynamics, even a small deviation from it (such as the power-Maxwell type of the Lagrangian) contains such solutions, and we have presented several classes of exact solutions with stealth NLE fields. The most important question left open is the existence of non-null stealth NLE field configuration on top of the black hole with nonvanishing electric or magnetic charge.

\begin{acknowledgments}
This research has been supported by the Croatian Science Foundation under Grant No.~8946. I thank Samuel Gralla and Ted Jacobson for several insightful remarks on force-free solutions.
\end{acknowledgments}

\bibliography{nlestealth}

\end{document}